\documentclass[aps,pre,onecolumn, amsfonts,showpacs, superscriptaddress]{revtex4-1}
\usepackage{epsfig,amsopn}
\usepackage{graphicx}
\usepackage{amsmath,amssymb}
\usepackage{natbib}
\usepackage{braket}
\newcommand{\lmax}{\lambda_{\max}}

\def\bea{\begin{eqnarray}}
\def\eea{\end{eqnarray}}

\def\nn{\nonumber}

\def\f{\frac}

\begin{document}

\title{Extreme value statistics of correlated random variables}

\author{Satya N. Majumdar} \affiliation{CNRS, LPTMS, Orsay, Paris Sud}

\author{Arnab Pal} \affiliation{Raman Research Institute, India}

\date{\today}

\begin{abstract} Extreme value statistics (EVS) concerns the study of the statistics of the maximum or the minimum 
of a set of random variables. This is an important problem for any time-series and has applications in climate, 
finance, sports, all the way to physics of disordered systems where one is interested in the statistics of the 
ground state energy. While the EVS of `uncorrelated' variables are well understood, little is known for strongly 
correlated random variables. Only recently this subject has gained much importance both in statistical physics and 
in probability theory. In this note, we will first review the classical EVS for uncorrelated variables and discuss 
few examples of correlated variables where analytical progress can be made.

\vskip 0.5cm

Lecture notes based on 2 lectures given by S.N. Majumdar during the training week in 
the GGI workshop `Advances in Nonequilibrium Statistical Mechanics: 
large deviations and long-range correlations, extreme value statistics, 
anomalous transport and long-range interactions' in Florence, Italy (May-2014).
The notes were prepared by Arnab Pal.

\end{abstract}

\maketitle

\section{Introduction}
\label{introduction}

Extreme events are ubiquitous in nature. They may be rare events but when they
occur, they may have devastating consequences and hence are rather
important from practical points of view. 
To name a few, different forms of natural calamities
such as earthquakes, tsunamis, extreme floods, large wildfire, the hottest and the coldest days,
stock market risks or 
large insurance losses in finance, new records in major sports events like Olympics 
are the typical examples of 
extreme events. There has been a major interest to study these 
events systematically using statistical methods and the field is known as Extreme Value Statistics
(EVS)~\cite{FT28,Gumbel,Gnedenko,Leadbetter}. This is a branch of statistics dealing with the extreme deviations from the mean/median 
of probability distributions (for a recent review on the subject
see~\cite{SM13} and references therein). The general theory sets out to assess the type of 
probability distributions generated by processes which are responsible for
these kind of highly unusual events. In recent years, it has been realized
that extreme statistics and rare events play an equally important role
in various physical/biological/financial contexts as well--for a few illustrative examples (by far not exhaustive) 
see~\cite{Derrida:81,Bouchaud:97,KM00,Satya:00,Dean:01,ADGR01,Raychowdhury,Satya:02,Racz:03,Ledoussal-Monthus,
MK03,Satya:04,airy,KM05,Satya:05,Bertin-Clusel,GMOR07,Sanjib,BM07,Krug07,RM07,CS07,CS107,Burkhardt,Satya:08,MB08,
MRKY08,MZ08,GL08,SMCR08,RMC09,LW09,GMS09,MCR10,SL10,MRZ10,MRZ110,NK11,RS11,FMS11,S12,SM13,SMCF13,Wergen13,DMRZ13,GMS14}.
A typical example 
can be found in disordered systems where the ground state energy, being
the minimum energy, plays the role of an extreme variable. In addition,
the dynamics at low temperatures in disordered systems are governed by the statistics of 
extremely low energy states. Hence the study of extremes and related quantities
is extremely important in the field of disordered systems~\cite{Derrida:81,Bouchaud:97,Satya:00,Dean:01,Ledoussal-Monthus,FB08,F09,FLR09,F10,MRZ10}. 
Another important physical system where extreme fluctuations play
an important role corresponds to fluctuating interfaces of the
Edwards-Wilkinson/Kardar-Parisi-Zhang varieties~\cite{Raychowdhury,Racz:03,Satya:04,airy,SM06,GMOR07,Burkhardt,RS09,RBKS11}.
Another exciting recent area concerns the distribution of the
largest eigenvalue in random matrices: the limiting distribution~\cite{TW94,TW96} and 
the large deviation probabilities~\cite{DM06,DM08,MV09} of the
largest eigenvalue and its various
applications (for a recent review on the largest eigenvalue of a random matrix, see~\cite{Satya:14}). 
Extreme value statistics also
appears in computer science problems such as in binary search trees
and the associated search algorithms~\cite{KM00,Satya:00,Satya:02,MK03,Satya:05}.

In the classical extreme value theory, one is concerned with the statistics
of the maximum (or minimum) of a set of {\em uncorrelated} random variables.
In contrast, in most of the physical systems mentioned above, the
underlying random variables are typically {\em correlated}. 
In recent years, there have been some advances in the understanding
of EVS of correlated variables. In these lectures, we will first review
the classical EVS of uncorrelated variables. Then we will discuss 
the EVS of {\em weakly} correlated random variables with some examples.
Finally few examples of {\em strongly} correlated random variables will
be discussed. 

It should be emphasised that this is not a review article, rather lecture notes based
only on two lectures. It would thus be impossible to cover, in only two lectures, all aspects of
this important and rapidly advancing field of extreme value statistics with
an enormous range of applications spanning across disciplines--from engineering sciences
all the way to physics. In these two lectures we will focus only on some basic and key
concepts. Consequently we will not attempt to provide an exhaustive list of references
in this broad subject and any inadvertent omission of relevant references is apologised
in advance.

\section{Extreme Value Statistics: Basic preliminaries}

In a given physical situation, one needs to first identify
the set of relevant random variables $\{x_1,x_2,\ldots, x_N\}$. For example, for fluctuating
one dimensional interfaces, the relevant random variables may denote
the heights of 
the interface at different space points. In disordered systems
such as spin glasses, $\{x_i\}$'s may denote the energy of different
spin configurations for a given sample of quenched disorder. Once
the random variables are identified, there are subsequently two basic steps
involved :
(i) to compute explicitly the joint distribution
$P(\{x_i\})$ of the relevant random variables (this is sometimes
very difficult to achieve) and (ii) suppose that we know the joint
distribution $P(\{x_i\})$ explicitly--then from this, how to compute
the distribution of some observables, such as the sample mean
or the sample maximum, defined as:
\bea
\text{Mean}~~~\bar{X}=\f{x_{1}+x_{2}+...+x_{N}}{N} \\
\text{Maximum}~~M=\text{max}~(x_{1},x_{2},...,x_{N})\,.
\eea

Particular simplifications occur for IID (independent and identically distributed)
random variables, where the
joint distribution $P(\{x_i\})$ factorises, i.e., 
$P(x_{1},x_{2},...,x_{N})=p(x_{1})p(x_{2})...p(x_{N})$, where
each variable is chosen from the same parent density 
$p(x)$. Knowing the parent distribution $p(x)$, one can then easily
compute the distributions, e.g., of $\bar{X}$ and of $M$. 

For example, let us first consider $\bar{X}$. 
One knows that irrespective of the choice of the parent distribution (with finite variance)
the PDF of the \textbf{mean} 
of the IID random variables tends to a Gaussian distribution for large N namely,
\bea
P(\bar{X},N)\xrightarrow{N\to \infty}\f{1}{\sqrt{2\pi \sigma^{2}/N}}e^{-\f{N}{2 \sigma^{2}}(\bar{X}-\mu)^{2}}
\eea
where $\mu,~\sigma^{2}$ are the mean and the variance of the parent distribution respectively. This is 
known as the \textbf{Central Limit Theorem} \cite{Feller:71} and this Gaussian form is universal. 
However, for correlated variables, one does not know, in general, how to compute
the distribution of $\bar{X}$ and consequently, there is no known universal law
for the limiting distribution of the mean of a set of correlated random variables. 

Similar question about universality also arises for the distribution of extremes, e.g.,
that of $M$. We will see below that, as in the case of the mean $\bar{X}$, there
exist universal limit laws for the distribution of the maximum $M$
for the case of IID variables. However, for {\em strongly} correlated
variables, the issue of universality is wide open.

Suppose that we know the joint distribution $P(\{x_i\})$ explicitly.
Then to compute the distribution of the maximum $M$, it is useful to
define the cumulative distribution of $M$ which can be easily expressed
in terms of the joint distribution
\bea
Q_{N}(x)&=&\text{Prob}[M\leq x,N]=\text{Prob}[x_{1}\leq x,~x_{2}\leq x,...,x_{N}\leq x]\\
&=&\int_{-\infty}^{x}
\int_{-\infty}^{x}...\int_{-\infty}^{x}dx_{1}dx_{2}...dx_{N}P(x_{1},x_{2},...,x_{N})
\eea
and the PDF of the maximum can be obtained by taking the derivative i.e. $P(M,N)=Q_{N}^{\prime}(x)$.

\section{Independent and Identical random variables}

For IID random variables, the joint PDF factorises and we get
\bea
Q_{N}(x)=[\int_{-\infty}^{x}dy~p(y)]^{N}=[1-\int_{x}^{\infty}dy~p(y)]^{N}\,.
\eea 
This is an exact formula for the cumulative distribution of the maximum for any $N$.
Evidently, $Q_N(x)$ depends explicitly on the parent distribution $p(x)$ for any
finite $N$. The question is: as in the CLT of the sum of random variables discussed
before, does any universality emerge for $Q_N(x)$ in the large $N$ limit? 
The answer is that indeed a form of universality emerges in the large $N$ limit, as
we summarize below (for a recent review, see~\cite{SM13}).

It turns out that in the scaling limit when $N$ is large, $x$ is large, with
a particular scaling combination (see below) fixed, $Q_N(x)$ approaches a limiting form:
\bea
Q_{N}(x)\xrightarrow[z=(x-a_{N})/b_{N}~\text{fixed}]{x \to \infty,~N\to \infty}F(\f{x-a_{N}}{b_{N}}) \\
\text{equivalently}~~~\lim_{N \to \infty}Q_{N}(a_{N}+b_{N}z)=F(z)
\eea
where $a_{N},~b_{N}$ are non-universal scaling factors that depend on the parent distribution $p(x)$,
but the scaling function $F(z)$ can only be of three possible varieties
$F_{I,II,III}(z)$ depending only on the large $x$ tail of the parent distribution
$p(x)$.   
This is known as the \textbf{Gnedenko}'s classical law of extremes (1943)~\cite{Gnedenko}.

\subsection{Parent distributions with a power law tail}

We consider the IID random variables whose parent distribution has a power law convergence 
$p(x)\sim x^{-(1+\alpha)}$ with $\alpha>0$. We denote the scaling function as $F_{1}(z)$ 
and this is found to be 
\bea
F_{1}(z)= \left\{
     \begin{array} {rl}
     {\rm e^{-z^{-\alpha}}} & ~ \text{for~} z \geq 0\\
     {\rm 0}  & ~ \text{for~}  z \leq 0 \\
     \end{array} \right.
\eea
The PDF is given by 
\bea
f_{1}(z)=\f{\alpha}{z^{\alpha+1}}e^{-z^{-\alpha}},~~~~~z\in[0,\infty)
\eea
Here one can identify $a_{N} \approx 0,~b_{N} \approx N^{1/\alpha}$ for large
$N$. This is the famous \textbf{Fr\'{e}chet} distribution.

\subsection{Parent distributions with a faster than power law tail}

We consider the parent distributions with tails which decay faster than power law but unbounded
such as $p(x)\sim e^{-x^{\delta}}$ with $\delta>0$. In this case, one finds the scaling function to be
\bea
F_{2}(z)=e^{-e^{-z}}
\eea
where as the PDF is given by
\bea
f_{2}(z)=e^{-z-e^{-z}},~~~~~z\in(-\infty,\infty)
\eea
This is the famous \textbf{Fisher-Tippett-Gumbel} distribution~\cite{FT28,Gumbel}. Here, 
for large $N$, one finds 
$a_{N} \approx (\ln N)^{1/\delta},~b_{N} \approx \f{1}{\delta}(\ln N)^{1/\delta-1}$.
Since $a_{N}$ is the typical value of the maximum and is defined by
$N\int_{a_{N}}p(y)dy=1$, the weight will be to the right of $a_{N}$.
In the following we will
compute the scaling functions for the parent distributions having 
the exponential and the Gaussian tails.
\subsubsection{$p(x)\sim e^{-x}$ for $x \geq 0$}
One finds that
\bea
Q_{N}(x)=[1-e^{-x}]^{N}=e^{N\log[1-e^{-x}]}\sim e^{-N e^{-x}}=e^{-e^{-(x-\log N)}}=F_{2}(z)
\eea 
with $z=x-\log N$ and one identifies $a_{N} \approx \ln N,~b_{N} \approx 1$
as for large $N$.
\subsubsection{$p(x)\sim e^{-x^{2}/2}$}
In the case of Gaussian tails, we find that
\bea
Q_{N}(x)\xrightarrow{x \to \infty,~N\to \infty}e^{-N\int_{x}^{\infty}p(y)dy}\sim 
e^{-N e^{-\f{x^{2}/2}{x\sqrt{\pi}}}}\sim e^{-e^{-(\f{x^{2}}{2}-\log N)}}=F_{2}(z)
\eea
which is peaked around $x=\sqrt{2 \log N}$. One again identifies this to be the
Fisher-Tippett-Gumbel distribution 
with $a_{N} \approx \sqrt{2 \ln N},~b_{N} \approx \f{1}{\sqrt{2 \ln N}}$
for large $N$.

\subsection{Parent distributions with a upper bounded support}

We now consider the parent distributions with the bounded tails
such as $p(x)\xrightarrow{x \to a} (a-x)^{\beta-1}$ with $\beta>0$. 
In this case 
\bea
F_{3}(z)= \left\{
     \begin{array} {rl}
     {\rm e^{-(-z)^{\beta}}} & ~ \text{for~} z \leq 0\\
     {\rm 1}  & ~ \text{for~}  z \geq 0 \\
     \end{array} \right.
\eea
The PDF is therefore given by
\bea
f_{3}(z)=\beta(-z)^{\beta-1}e^{-(-z)^{\beta}},~~~~~z\in(-\infty,0]
\eea
This is well-known as the \textbf{Weibull} distribution.

Let us now summarize the results for the limiting distribution of the maximum
for IID random variables in the following table.

\begin{center}
    \begin{tabular}{ | l | l | l | p{3cm} |}
    \hline
    parent distribution $p(x)$ & scaling function $F(z)$ & PDF of maximum $f(z)$ & Nomenclature\\ \hline
    $x^{-(1+\alpha)};\,\, \alpha>0$ & $e^{-z^{-\alpha}}\theta(z)$    
& $\f{\alpha}{z^{\alpha+1}}e^{-z^{-\alpha}};\,\, z>0 $ &  \textbf{Fr\'{e}chet}  \\ \hline
    $e^{-x^{\delta}}$ & $e^{-e^{-z}}$                  
& $e^{-z-e^{-z}}$                            &  \textbf{Fisher-Tippett-Gumbel}       \\ \hline
    $(a-x)^{\beta-1};\,\, \beta>0$ & $e^{-(-z)^{\beta}}
\theta(-z)+\theta(z)$  & $\beta(-z)^{\beta-1}e^{-(-z)^{\beta}};\,\,\, z<0$ &   \textbf{Weibull} \\
    \hline
    \end{tabular}
\end{center}

So far, we discussed about the limiting laws in the limit of large sample size $N$.
It turns out however that the convergence to these limiting laws is extremely
slow and in simulations and experiments, it is very hard to see these limiting distributions~\cite{GMOR08,astro}.
A renormalization group treatment has been recently developed that describes
how these EVS distributions for IID variables approach their limiting fixed
point distributions. Interested readers may discuss Refs.~\cite{GMOR08,GMORD10,BG10,ABA12,CCEF12}.

\subsection{Order statistics}

An interesting generalisation of the statistics of the global maximum corresponds to studying
the statistics of successive maxima, known as the `order' statistics~\cite{ABN,ND}.
For a recent review on order statistics, see ~\cite{SM13}. Consider again the set of IID random variables
$\{x_{1},x_{2},...,x_{N}\}$ and arrange them in decreasing
order of their values.
So, if we denote them by $M_{k,N}$ where $k$ is the order and $N$ is the number 
of variables, then
\bea
M_{1,N}&=&\text{max~}(x_{1},x_{2},...,x_{N})  \nn \\
M_{2,N}&=&\text{secondmax~}(x_{1},x_{2},...,x_{N}) \nn \\
\dots \nn \\
M_{k,N}&=&\text{k-thmax~}(x_{1},x_{2},...,x_{N})  \nn \\
\dots \nn \\
M_{N,N}&=&\text{min~}(x_{1},x_{2},...,x_{N})
\eea
and henceforth by definition 
$M_{1,N}>M_{2,N}>...>M_{N,N}$. It is of interest to study
the statistics of the k-th maximum $M_{k,N}$ and the statistics of the 
gap defined by $d_{k,N}=M_{k,N}-M_{k+1,N}$. We consider the parent distribution
to be $p(x)$. Then we define the upward and the downward 
cumulative distributions respectively
\bea
p_{>}(x)=\int_{x}^{\infty}p(y)dy \\
p_{<}(x)=\int_{-\infty}^{x}p(y)dy
\eea
Then the cumulative probability $Q_{k,N}(x)$
that the k-th maximum stays below $x$ is given by
\bea
Q_{k,N}(x)=\text{Prob}[M_{k,N} \leq x]=\sum_{m=0}^{k-1}{N \choose m}~[p_{>}(x)]^{m}[p_{<}(x)]^{N-m}
\eea
where $m$ is the no. of maximums above $x$. The exact PDF is then given by
\bea
P_{k,N}=\text{Prob}[M_{k,N}=x]=N {N-1 \choose k-1}~p(x)~[p_{<}(x)]^{N-k}[p_{>}(x)]^{k-1}
\eea
The gap distribution is also very straightforward to obtain from here
\bea
P_{k,N}(d)=\text{Prob}[d_{k,N}=d]=N(N-1){N-2 \choose k-1} \int_{-\infty}^{\infty}dx~p(x)p(x-d)[p_{<}(x)]^{N-k-1}[p_{>}(x)]^{k-1}
\eea
As for the cumulative distribution for order statistics, we can again do the same analysis as before and 
extract the leading scaling behavior 
\bea
Q_{k,N}(x)=\text{Prob}[M_{k,N} \leq x]\xrightarrow[z=(x-a_{N})/b_{N}~\text{fixed}]{x \to \infty,~N\to \infty}G_{k}(\f{x-a_{N}}{b_{N}})
\eea
where the scaling function $G_{k}(z)$ is given by 
\bea
G_{k}(z)=F_\mu(z) \sum_{j=0}^{k-1}\f{[-\ln F_\mu(z)]^{j}}{j!}= \frac{1}{\Gamma(k)}\,
\int_{-\ln F_\mu(z)}^{\infty} e^{-t}\, t^{k-1}\, dt
\eea
where $F_{\mu}(z)$, with $\mu=1,2,3$,  denote respectively the Fr\'echet, Fisher-Tippett-Gumbel
and Weibull scaling functions for the global maximum discussed in the previous subsection. One can
clearly see that for $k=1$ (global maximum), indeed, $G_1(z)= F_\mu (z)$. 
As an example,
if we choose the parent distribution $p(x)$ from the subclass B where $F_2(z)=e^{-e^{-z}}$ 
is the Fisher-Tippett-Gumbel distribution, we find
$G_{k}(z)=e^{-e^{-z}}\sum_{j=0}^{k-1}\f{e^{-jz}}{j!}$ and the PDF is $G_{k}^{\prime}(z)=\f{e^{-kz-e^{-z}}}{(k-1)!}$.
This is often known as the generalized Fisher-Tippett-Gumbel law. One can also derive
exactly the limiting scaling distribution of the $k$-th gap, $d_{k,N}= M_{k,N}-M_{k+1,N}$ (see Ref.~\cite{SM13}
for details).

\section{correlated random variables}
In this section we study the extreme statistics of the correlated random variables. 
In the first subsection we revisit 
the random variables where the correlation is weak and then we study the `strongly' interacting random variables.
There is no general framework to study the statistics of the correlated random variables.
However, we show in the following that for the weakly correlated variables, one can provide a
a rather general renormalization group type of argument to study the extreme statistics and
that should hold in general. 
Nevertheless, this argument does not work when the variables are strongly
correlated and one has to study case by case in different models to
gain an insight.

\subsection{Weakly correlated random variables} 

Suppose that we have a set of random variables that are not independent, but correlated
such that the connected part of the correlation function decays fast (say exponentially)
over a certain finite correlation length $\zeta$
\bea
C_{i,j}=\langle x_{i}x_{j} \rangle-\langle x_{i}\rangle\langle x_{j}\rangle\sim~e^{-|i-j|/\zeta}
\eea 
Clearly, when the two variables are separated over a length scale larger than $\zeta$, i.e.,
when $|i-j|>>\zeta$, then they essentially get uncorrelated. 
Now weak correlation implies that $\zeta<<N$, where
$N$ is the total size of the sample.

For such weakly correlated variables one can construct a heuristic argument
to study the extreme statistics~\cite{airy}, as we describe now.
Consider $N'=\zeta<< N$ and break the system into identical blocks
each of size $\zeta$.
There
are thus $N/\zeta$ no. of blocks. 
While the random variables inside each box are still strongly correlated,
the variables that belong to different boxes are approximately uncorrelated.
So, each of these boxes are 
non-interacting. 
Now, for each box $i$, let $y_i$ denote the `local maximum', i.e., the maximum
of all the $x$-variables belonging to the $i$-th block, where $i=1,2,\ldots, N'=N/\zeta$.
By our approximation, $y_i$'s are thus essentially {\em uncorrelated}.
So, we have
\bea
M=\text{Max}[x_{1},x_{2},...,x_{N}]=\text{Max}[y_{1},y_{2},...,y_{N'}]
\eea
So, in principle if one knows the PDF of $y$, then this problem is essentially
reduced to calculating the maximum of $N'$ uncorrelated random variables
$\{y_1,y_2,\ldots, y_{N'}\}$, which has already been discussed before.
So, we know that depending on the tail of $p(y)$, the limiting distribution
of $M$ of $N$ weakly correlated variables will, for sure, belong to
one of three (Fr\'echet, Fisher-Tippett-Gumbel or the Weibull class)
limiting extreme distributions of IID random variables.
To decide the tail of $p(y)$, of course one needs to solve a {\em strongly}
correlated problem since inside each block the variables are strongly correlated.
However, one can often guess the tail of $p(y)$ without really solving for
the full pdf of $p(y)$ and then one knows, for sure, to which class the
distribution of the maximum belongs to. As a concrete example
of this procedure for weakly correlated variables, we discuss in the next section the Ornstein-Uhlenbeck
stochastic process where one can compute the EVS exactly and demonstrate that
indeed this heuristic renormalization group argument works very well.
To summarize, the problem of EVS of weakly correlated random variables 
basically reduces to IID variables with an effective number $N'=N/\zeta$ where
$\zeta$ is the correlation length. So, the real challenge is to compute
the EVS of strongly correlated variables where $\zeta\ge O(N)$, to which
we now turn to below.

\subsection{Strongly correlated random variables}

Strongly correlated means that $\zeta>>N$ i.e. a  
correlation prevails over the whole system and the idea of block 
spins will no longer hold. A general theory for calculating the
EVS, such as in case
of IID or weakly correlated variables, is currently lacking
for such strongly correlated variables. In absence of a general theory,
one tries to study different exactly solvable special cases in the hope
of gaining some general insights. There are examples, though their numbers
are unfortunately few, where the EVS of a strongly correlated system can be computed
exactly. In the rest of the lecture, we will discuss a few examples of
exactly solvable cases.

As a first example of a strongly correlated system, we will consider the one dimensional
Brownian motion, for which the distribution of maximum can be computed explicitly.
Next, we will discuss another process called the Ornstein-Uhlenbeck (OU) process, which
represents the noisy motion of a classical particle in a harmonic well. We will see
that the OU process represents a weakly correlated system, for which one can
compute the distribution of the maximum explicitly, demonstrating the power
of heuristic argument presented in the previous subsection for weakly
correlated variables.  
Then we will 
present the problem of the maximum height distribution of a fluctuating $(1+1)$-dimensional 
interface in its stationary state.
Finally, we will discuss the statistics of the largest eigenvalue for Gaussian random matrices.

\subsubsection{One dimensional Brownian motion}

We consider the case where the random variables $\{x_i\}$ represent the positions
of a one dimensional random walker at discrete time-step $i$, starting from $x_0=0$.
We are interested in the maximum position of the walker up to step $n$. Even though
the discrete problem can be solved explicitly, for simplicity we will consider
below the continuous-time version of the random walk, i.e., a one dimensional
Brownian motion whose position $x(\tau)$ evolves via the stochastic Langevin equation
\bea
\f{dx}{d\tau}=\eta(\tau)
\eea
starting from $x(0)=0$ and
$\eta$ represents a Gaussian white noise with $\langle \eta(\tau) \rangle=0$ 
and $\langle \eta(\tau)\eta(\tau') \rangle=2D\delta(\tau-\tau')$. 
We are interested in the
PDF of the maximum $M(t)$ of this Brownian motion $x(\tau)$ over the time window $[0,t]$:
$M(t) = \max_{0\le \tau\le t}\left[x(\tau)\right]$.

To proceed, we first note that $x(\tau)= \int_0^{\tau} \eta(s)\, ds$ and hence, one can easily
compute the mean and the correlator of the process $x(\tau)$: 
$\langle x(\tau) \rangle=0$ and $\langle x(\tau)x(\tau') \rangle=2D~\text{min}(\tau,\tau')$.
Thus the variables $x(\tau)$ at different times are strongly correlated. The
correlation function does not decay and persists over the whole sample size $\tau\in [0,t]$.

It is again very useful to define the 
cumulative distribution of the maximum  
\bea
Q(z,t)&=&\text{Prob}[M(t)\leq z] \nn \\
&=&\text{Prob}[x(\tau)\leq z,~0 \leq \tau \leq t] \nn \\
\text{where~}M(t)&=&\text{max}_{0 \leq \tau \leq t}[x(\tau)]
\eea
To compute $Q(z,t)$, we note that it just represents the
probability that the Brownian particle, starting at $x(0)=0$, stays below
the level $z$ up to time $t$. Let $P(x,t|z)$ denote the
probability density for the particle to reach $x$ at time $t$, while staying below
the level $z$ during $[0,t]$. It is then easy to see that $P(x,t|z)$ satisfies
the diffusion equation in the semi-infinite domain $x\in [-\infty, z]$ with
the following boundary and initial conditions:
\bea
\f{\partial P}{\partial t}&=&D\f{\partial^{2}P}{\partial x^{2}} \nn \\
\text{with~}P(x,0|z)&=&\delta(x) \nn \\
\text{and~}P(x\to -\infty,t|z)&=&0,~P(x=z,t|z)=0\,.
\eea
The absorbing boundary condition $P(x=z,t|z)=0$ at $x=z$ guarantees that
the particle does not cross the level at $x=z$. 
The solution is trivial and can be solved by the method of images~\cite{redner,satya_review10,BMS13}
\bea
P(x,t)=\f{1}{\sqrt{4\pi Dt}}[e^{-\f{x^{2}}{4Dt}}-e^{-\f{(x-2z)^{2}}{4Dt}}]
\eea
Therefore the cumulative and the PDF of the maximum are given by
\bea
Q(z,t)=\int_{-\infty}^{z}dx~P(x,t)={\rm erf}\left(\f{z}{\sqrt{4Dt}}\right);\,\, {\rm where}\,\, 
{\rm erf}(z)= \frac{2}{\sqrt{\pi}}\,\int_0^z e^{-u^2}\,du \\
{\rm Prob.}(M=z,t)=Q^{\prime}(z)=\f{1}{\sqrt{\pi Dt}}e^{-\f{z^{2}}{4Dt}}\theta(z)
\eea
One easily finds that the mean $\langle M(t) \rangle = \frac{2}{\sqrt{\pi}}\,\sqrt{D\,t}$.
This thus represents, perhaps, the simplest example of a strongly correlated system
for which one can compute the distribution of the maximum exactly.

For more recent works on the global maximum and the order/gap statistics of discrete-time random walks,
L\'evy flights, $1/f^{\alpha}$ signals see Refs.~\cite{CM05,satya_review10,BMS13,MOR11,SM12,FM12,MMS13,PCMS13}. The order/gap
statistics for one dimensional branching Brownian motion has also been studied
recently and a number of analytical results are available~\cite{BD1,BD2,RMS14}.

\subsubsection{Ornstein-Uhlenbeck (OU) process}
Consider a Brownian particle in a harmonic potential governed by the 
following equation
\bea
\f{dx}{d\tau}=-\mu x+\eta(\tau)
\label{ou1}
\eea 
where, as before, $\eta(\tau)$ is a Gaussian white noise with zero mean and
is delta correlated, $\langle \eta(\tau)\eta(\tau')\rangle = 2\,D\, \delta(\tau-\tau')$.
Assuming that the particle starts at the origin $x(0)=0$, the correlation
function between two times can be trivially computed and is given by
\bea
C(t_{1},t_{2})=\langle x(t_{1})x(t_{2}) \rangle=\f{D}{\mu}[e^{-\mu|t_{1}-t_{2}|}-e^{-\mu(t_{1}+t_{2})}]
\eea
Clearly, in the limit $\mu\to 0$, the correlation function reduces to the Brownian limit:
$C(t_1,t_2) \to 2\, D\, {\rm min}(t_1,t_2)$ as $\mu\to 0$ and the system becomes strongly correlated.
In contrast, for nonzero $\mu>0$, the correlation function,
at large times when $t_{1},t_{2}>1/\mu$, decays exponentially with the time-difference:
$C(t_{1},t_{2})=\f{1}{2\mu}e^{-\mu|t_{1}-t_{2}|}$
with a correlation length $\zeta=1/\mu$. From our arguments before about weakly
correlated random variables, we would then expect to get the limiting Fisher-Tippett-Gumbel
distribution for $\mu>0$. We demonstrate below briefly how actually this
Fisher-Tippett-Gumbel distribution emerges by actually solving exactly the EVS problem
for the OU process. 

As before, let $Q(z,t)$ denote the cumulative distribution of the maximum
$M(t)$ of the OU process in the time interval $[0,t]$. The particle
starts at the origin $x(0)=0$ and evolves via Eq. (\ref{ou1}).
Let $P(x,t|z)$ denote the probability density for the particle to
arrive at $x$ and time $t$, while staying below the level $z$.
This restricted propagator satisfies the Fokker-Planck equation
in the domain $x\in [-\infty,z]$
\bea
\f{\partial P}{\partial t}&=&D\f{\partial^{2}P}{\partial x^{2}}+\mu \f{\partial}{\partial x}[xP]
\label{fp.ou}
\eea
with the initial condition: $P(x,0|z)= \delta(x)$ and the boundary conditions:
$P(x,t|z)=0$ as $x\to -\infty$ and also the absorbing condition at level $z$: $P(x=z,t|z)=0$
for all $t$. For simplicity, we will set $D=1/2$.
We note that, unlike in the Brownian case ($\mu=0$), for $\mu>0$ we can
no longer use the method of images due to the presence of the potential. 
However, one can solve this equation by the eigenfunction expansion and the solution
can be expressed as 
\bea
P(x,t|z)=\sum_{\lambda} a_{\lambda}\, e^{-\lambda t}\, D_{\lambda/\mu}(-\sqrt{2\mu}x)\, e^{-\mu\, x^2/2}
\label{sol1_ou}
\eea
where  $D_{p}(z)$ is the 
parabolic cylinder function which satisfies the second order ordinary differential
equation: $D_p''(z)+ (p+1/2-z^2/4)\,D_p(z)=0$ (out of the two linearly independent
solutions, we choose the one that vanishes as $z\to \infty$). The absorbing boundary condition
$P(x=z,t|z)=0$ induces the boundary condition on the eigenfunction:  
$D_{\lambda/\mu}(-\sqrt{2\mu}z)=0$, which then fixes the eigenvalues $\lambda$'s.
One gets the spectrum of eigenvalues and they are necessarily positive.
At large times $t$, the summation in Eq. (\ref{sol1_ou}) will be dominated
by the term involving the smallest eigenvalue ${\lambda}_0 (z)$, which evidently
depends on $z$. For arbitrary $z$, it is difficult to solve 
$D_{\lambda/\mu}(-\sqrt{2\mu}z)=0$ and determine the smallest eigenvalue
$\lambda_0(z)$. However, for large $z$, one can make progress by perturbation
theory and one can show that to leading order for large $z$,
\bea
\lambda_0(z)\xrightarrow{z\to \infty}\f{2}{\sqrt{\pi}}\,\mu^{3/2}\,z\,e^{-\mu z^{2}}.
\label{l0}
\eea
Consequently, for large $t$ and large $z$,
\bea
Q(z,t)\sim e^{-\lambda_0(z)\,t}&\sim& e^{-e^{-\mu z^{2}+\log (\f{2 t \mu^{3/2}z}{\sqrt{\pi}})}} \nn \\
&\rightarrow&F_{2}\left(\sqrt{4\,\mu\, \ln t}\left(z-\frac{1}{\sqrt{\mu}}\, \sqrt{\ln t}\right)\right)
\label{ou_gumbel}
\eea
where $F_{2}(y)=\exp[-\exp[-y]]$ is the Fisher-Tippett-Gumbel distribution. 
As a result, for $\mu>0$, the average value of the maximum grows very slowly for large $t$ as,
$\langle M(t)\rangle \sim \frac{1}{\sqrt{\mu}}\, \sqrt{\ln t}$, while its 
width around the mean decreases as $\sim 1/\sqrt{\ln t}$.

Indeed for $\mu>0$, a full analysis of the mean value of the maximum $\langle M(t)\rangle$ for all $t$
shows that initially it grows as $\sqrt{t}$ (for $t<< 1/\mu$) where it does not feel the
confining potential and hence behaves as a Brownian motion. But for $t>> 1/\mu$, the particle
feels the potential and the mean maximum crosses over to a slower growth as $\sqrt{\ln t}$
\bea
\langle M(t) \rangle \sim \left\{
     \begin{array} {rl}
     {\rm \sqrt{t}}     & ~ \text{for~} t << 1/\mu \\
     {\rm \sqrt{\ln t}} & ~ \text{for~} t >> 1/\mu \\
     \end{array} \right.
\eea

\subsubsection{Fluctuating interfaces in 1D}
The most well studied model of a
fluctuating (1 + 1)-dimensional surfaces is the so called Kardar-Parisi-Zhang (KPZ)
equation that describes the time evolution of the height $H(x,t)$ of an interface 
growing over a linear substrate of size $L$ via the stochastic partial
differential equation
\bea
\f{\partial H}{\partial t}=\f{\partial^{2}H}{\partial x^{2}}+\lambda 
\left(\f{\partial H}{\partial x}\right)^{2}+\eta(x,t)
\eea
where $\eta(x,t)$ is a Gaussian white noise with zero mean and a 
correlator $\langle \eta(x,t)\eta(x^{\prime},t^{\prime}) \rangle=2\delta(x-x^{\prime})\delta(t-t^{\prime})$. 
For $\lambda=0$, the equation becomes linear and 
is known as the Edwards-Wilkinson (EW) equation. 
For nice reviews on fluctuating interfaces, see e.g.~\cite{HZ95,Krug97}.
The height is usually measured 
relative to the spatially averaged height i.e.
\bea
h(x,t)=H(x,t)-\f{1}{L}\int_{0}^{L}H(y,t)dy \\
\text{with~}\int_{0}^{L}h(x,t)dx=0
\eea
It can be shown that the joint PDF of the relative height field
$P(\{h\},t)$ reaches a steady state as $t \to \infty$ in a finite system of size $L$.
Also the height variables are strongly correlated in the stationary state.
Again in the context of the EVS, a quantity that has created some interests recently
is the PDF of the maximum relative height in the stationary state, i.e. $P(h_{m},L)$
where
\bea
h_{m} = \lim_{t \to \infty} \text{max}_x [{h(x, t)}, 0 \leq x \leq L] .
\eea
This is an important physical quantity that measures the extreme fluctuations 
of the interface heights~\cite{Raychowdhury,Satya:04}. We assume
that initially the height profile is flat. 
As time evolves, the heights of the interfaces at different spatial points
grow more and more correlated. The correlation length typically grows as
$\zeta\sim t^{1/z}$ where $z$ is the dynamical exponent ($z=3/2$ for KPZ and
$z=2$ for EW interfaces). For $t<< L^{z}$, the interface is in the `growing' regime
where again the height variables are weakly correlated since $\zeta\sim t^{1/z}<<L$.
In contrast, for $t>> L^z$, the system approaches a `stationary' regime
where the correlation length $\zeta$ approaches the system size and hence
the heights become strongly correlated variables.

Following our general argument for weakly correlated variables, we would then
expect that in the growing regime the maximal relative height, appropriately centred
and scaled, should have the Fisher-Tippett-Gumbel distribution.
In contrast, in the stationary regime, the height variables are strongly
correlated and the maximal relative height $h_m$ should have a different
distribution. This distribution was first computed numerically in ~\cite{Raychowdhury}
and then it was computed analytically in Refs.~\cite{Satya:04,airy}. This then
presents one of the rare solvable cases for the EVS of strongly correlated
random variables. Below, we briefly outline the derivation of this distribution. 

The joint PDF of the relative heights in the stationary state can be written 
putting all the constraints together~\cite{Satya:04,airy}
\bea
P_{st}[\{h\}]=C(L)e^{-\f{1}{2}\int_{0}^{L}(\partial_{x}h)^{2}~dx} \times \delta \Big[h(0)-h(L)\Big] \times \delta\Big[\int_{0}^{L}h(x,t)dx\Big]
\label{stat_measure}
\eea
where $C(L)=\sqrt{2\pi L^{3}}$ is the normalization constant and can be obtained integrating over all 
the heights. Note that this stationary measure of the relative heights 
is independent of the coefficient $\lambda$ of the nonlinear term in the KPZ equation, implying
that the stationary measure of the KPZ and the EW interface is the same
in $(1+1)$-dimension. But this is a special property only in $(1+1)$-dimension.
The stationary measure indicates that the interface locally behaves as a Brownian motion 
in space~\cite{HZ95,Krug97}. 
For an interface with periodic boundary condition, one would then
have a Brownian bridge in space.
However, 
it turns out that the constraint $\int_0^L h(x,t)\, dx=0$ (the zero mode
being identically zero), as shown explicitly
by the delta function in  Eq. (\ref{stat_measure}), plays an important role for
the statistics of the maximal relative height~\cite{Satya:04}. It shows that
the stationary measure of the relative heights actually corresponds to
a Brownian bridge, but with a global constraint that the area under
the bridge is strictly zero~\cite{Satya:04,airy}. This is an important
fact that plays a
crucial role for the extreme statistics of relative heights~\cite{Satya:04,airy}.

We define the cumulative distribution of the maximum relative height
$Q(z,L)=\text{Prob}[h_{m}\leq z]$. The PDF of the maximum relative height is 
then $P(z,L)=Q^{\prime}(z,L)$. 
Clearly $Q(z,L)$ is also the probability that the heights
at all points in $[0, L]$ are less than $z$ and can be formally
written in terms of the path integral~\cite{Satya:04,airy}
\bea
Q(z,L)=C(L)\int_{-\infty}^{z}du\int_{h(0)=u}^{h(L)=u}\mathcal{D}h(x)e^{-\f{1}{2}\int_{0}^{L}(\partial_{x}h)^{2}~dx} \nn \\
\times  \delta\Big[\int_{0}^{L}h(x,t)dx\Big]I(z,L)
\eea
where $I(z,L)=\prod_{x=0}^{L}\theta[z-h(x)]$ is an indicator function 
which is $1$ if all the heights are less than $z$ and zero
otherwise. Using the path integral technique, this 
integral can be computed exactly (for details see ~\cite{Satya:04,airy}). 
It was found that the PDF of $h_m$ has the scaling form for all $L$
\bea
P(h_{m},L)=\f{1}{\sqrt{L}}f\left(\f{h_{m}}{\sqrt{L}}\right)
\eea
where the scaling function can be computed explicitly
as~\cite{Satya:04,airy}
\bea
f(x)=\f{2\sqrt{6}}{x^{10/3}}\sum_{k=1}^{\infty}e^{-\f{b_{k}}{x^{2}}}\,b_{k}^{2/3}\,U\left(
-\f{5}{6},\f{4}{3},\f{b_{k}}{x^{2}}\right)
\eea
where $U(a,b,y)$ is the confluent hypergeometric function 
and $b_{k}=\f{2}{27}\alpha_{k}^{3}$, where $\alpha_k$'s are
the absolute values of the zeros of Airy function: ${\rm Ai}(-\alpha_k)=0$.
It is easy to obtain the small $x$ behavior of $x$ since 
only the $k=1$ term dominates as $x \to 0$.
Using $U(a,b,y)\sim y^{-a}$ for large $y$, we get as $x \to 0$,
\bea
f(x)\to \f{8}{81}\alpha_{1}^{9/2}\,x^{-5}\,\exp\left[-\f{2 \alpha_{1}^{3}}{27x^{2}}\right]
\eea
The asymptotic behavior of $f(x)$ at large $x$ can be obtained as 
\bea
f(x)\xrightarrow{x\to \infty}e^{-6 x^{2}}
\eea
It turns out, rather interestingly, that this same function has appeared before
in several different problems in computer science and probability theory and
is known in the literature as the Airy distribution function (for a review on
this function and its appearances in different contexts see Ref.~\cite{airy}
and references therein). 

The path integral technique mentioned above for computing the
maximal relative height distribution of the EW/KPZ stationary interfaces
have subsequently been generalised to more complex interfaces~\cite{SM06,GMOR07,Burkhardt,RS09,RBKS11}.

\subsubsection{Largest eigenvalue in the random matrix theory}

Another beautiful solvable example of the extremal statistics of strongly correlated variables
can be found in the random matrices~\cite{TW94,TW96}, see~\cite{Satya:14} for a recent review
from physics perspectives.
Let us consider a $N\times N$ Gaussian random matrices 
with real symmetric, complex Hermitian,
or quaternionic self-dual entries
$X_{i,j}$ distributed via the joint Gaussian law: 
\bea
\text{Pr}[\{X_{i,j}\}]\propto \exp[-\f{\beta}{2}N\text{Tr}(X^{2})],
\eea
where $\beta$ is the Dyson index. The distribution is invariant respectively
under orthogonal, unitary and symplectic rotations giving rise
to the three classical ensembles: Gaussian orthogonal ensemble (GOE),
Gaussian unitary ensemble (GUE) and Gaussian symplectic ensemble (GSE).
The quantized values of $\beta$ are respectively $\beta=1~$(GOE),
$\beta=2~$(GUE) and $\beta=4~$(GSE). The eigenvalues and eigenvectors are
random and their joint distribution decouple. Integrating out the eigenvectors,
here we focus only on the statistics of $N$ eigenvalues $\lambda_{1},
~\lambda_{2},...,~\lambda_{N}$ which are all real. The joint PDF of these eigenvalues
is given by the classical result
\bea
P_{\text{joint}}(\lambda_{1},~\lambda_{2},...,~\lambda_{N})=B_{N}(\beta)\exp \bigg[ -
\f{\beta}{2}N\sum_{i=1}^{N} \lambda_{i}^{2} \bigg] \prod_{i<j}|\lambda_{i}-\lambda_{j}|^{\beta},
\eea
where $B_{N}(\beta)$ is the normalization constant. For convenience, we rewrite the statistical weight as
\bea\label{joint_pdf}
P_{\text{joint}}(\lambda_{1},~\lambda_{2},...,~\lambda_{N})=B_{N}(\beta)
\exp \Bigg[ -\beta \Big( \f{N}{2}\sum_{i=1}^{N} \lambda_{i}^{2}-\f{1}{2} \sum_{i \neq j}\ln |\lambda_{i}-\lambda_{j}| \Big) \Bigg],
\eea

Hence, this joint law can be interpreted as a Gibbs-Boltzmann
measure (Dyson, 1962),
$P_{\rm joint}(\{\lambda_i\})\propto \exp\left[-\beta\,
E\left(\{\lambda_i\}\right)\right]$,
of an interacting gas of charged particles on a line where $\lambda_i$
denotes
the position of the $i$-th charge and $\beta$ plays the role of the
inverse temperature. The energy $E\left(\{\lambda_i\}\right)$ has
two parts: each pair of charges repel
each other via a $2$-d Coulomb (logarithmic)
repulsion (even though the charges are confined on the $1$-d
real line)
and
and
each charge is subject to an external confining parabolic potential.
Note that while $\beta = 1$, $2$ and $4$ correspond to the three
classical rotationally invariant Gaussian ensembles, it is
possible to associate a   
matrix model
to (\ref{joint_pdf}) for any value of $\beta > 0$ (namely tridiagonal
random matrices introduced by Dimitriu and Edelman in 2002).
Here we focus on
the largest eigenvalue $\lmax = \max_{1 \leq i \leq N} \lambda_i$: what
can be said about its fluctuations, in particular when $N$ is large ?
This
is a non trivial question as the interaction term, $\propto
|\lambda_i - \lambda_j|^\beta$, renders inapplicable the classical results
of extreme value statistics for
IID random variables.

The two terms in the energy of the Coulomb gas in (\ref{joint_pdf}), the
pairwise Coulomb repulsion and the external harmonic potential, compete
with each other. While the former tends to spread the charges apart, the  
later tends to confine the charges near the origin. As a result of this
competition, the system of charges settle down into an equilibrium
configuration on an average
and the average density of the charges is given by
\bea
\rho_{N}(\lambda)=\f{1}{N}\Big \langle
\sum_{i=1}^{N}\delta(\lambda-\lambda_{i}) \Big \rangle
\eea
where the angular brackets denote an average over with respect to the joint
PDF in Eq. (\ref{joint_pdf}). For such Gaussian matrices (\ref{joint_pdf}), it is well
known (Wigner, Dyson) that as $N\to \infty$,
the average density approaches an $N$-independent limiting form which
has a semi-circular shape
on the compact support $[-\sqrt{2}, + \sqrt{2}]$
\bea
\lim_{N\to \infty} \rho_{N}(\lambda)=\tilde{\rho}_{sc}(\lambda)=\f{1}{\pi}\sqrt{2-\pi^{2}}
\eea
where $\tilde{\rho}_{sc}(\lambda)$ is called the Wigner semi-circular law. Hence our first observation is 
that the maximum eigenvalue resides near the upper edge of the Wigner semi-circle:
\bea
\lim_{N\to \infty} \langle \lambda_{\text{max}} \rangle=\sqrt{2}
\eea
However, for large but finite $N$, $\lambda_{\text{max}}$ will
fluctuate from sample to sample and the interesting thing would be
to compute the cumulative distribution which is 
\bea
Q_{N}(w)=\text{Prob}[\lambda_{\text{max}}<w]
\eea
which can be written as a ratio of two partition functions
\bea
Q_{N}(w)&=&\f{Z_{N}(w)}{Z_{N}(w\to \infty)}, \\
Z_{N}(w)&=&\int_{-\infty}^{w}d\lambda_{1}...\int_{-\infty}^{w}d\lambda_{N}\exp \Bigg[ 
-\beta \Big( \f{N}{2}\sum_{i=1}^{N} \lambda_{i}^{2}-\f{1}{2} \sum_{i \neq j}\ln |\lambda_{i}-\lambda_{j}| \Big) \Bigg] 
\eea
where the partition function describes a 2-d Coulomb gas, confined 
on a 1-d line and subject to a harmonic potential, in the presence of a hard wall at $z$.
The study of this ratio of two partition functions reveals the existence of 
two distinct scales correspond to $(i)$ \textbf{typical} fluctuations of the 
top eigenvalue, where $\lambda_{\text{max}}=\mathcal{O}(N^{-2/3})$
and $(ii)$ \textbf{atypical} large fluctuations, where $\lambda_{\text{max}}=\mathcal{O}(1)$.
It can be shown that the typical fluctuations are governed by
\bea
\lambda_{\text{max}}=\sqrt{2}+\f{1}{\sqrt{2}}N^{-2/3}\chi_{\beta}
\eea
where $\chi_{\beta}$ is an $N$-independent random variable. Its cumulative distribution,
$\mathcal{F}_{\beta}=\text{Prob}[\chi_{\beta}\leq x],$ is known as the 
$\beta$-Tracy-Widom (TW) distribution which is known only for $\beta=1,~2,~4$~\cite{TW94,TW96}.
For arbitrary $\beta>0$, it can be shown that this PDF has asymmetric 
non-Gaussian tails~\cite{TW94,TW96},
\bea
\mathcal{F}_{\beta}^{\prime}(x) \approx \left\{
     \begin{array} {rl}
     {\rm \exp~\Big[-\f{\beta}{24}|x|^{3}\Big]}, & ~~~x \to  -\infty\\
         \\
     {\rm \exp~\Big[-\f{2\beta}{3}x^{3/2}\Big]}  & ~~~x \to  +\infty \\
     \end{array} \right.
\eea
While the TW density describes the probability of typical fluctuations 
of $\lambda_{\text{max}}$ around
its mean $\langle \lambda_{\text{max}} \rangle=\sqrt{2}$ on a small 
scale of $\mathcal{O}(N^{-2/3})$, it does not describe atypically
the large fluctuations, e.g. of order $\mathcal{O}(1)$ around its mean.
The probability of atypical large
fluctuations, to leading order for large $N$, is described by two 
large deviations (or rate)
functions $\Phi_{-}(w)$ (for fluctuations to the left of the mean) 
and $\Phi_{+}(w)$ (for fluctuations to
the right of the mean), which were computed respectively
in Ref.~\cite{DM06,DM08} and in Ref.~\cite{MV09}.
The left and the right tails are equivalent to the physical
situations of pushed and pulled Coulomb gas against the wall (see
\cite{Satya:14} for a review).
We now present the asymptotic results 
which are known for the PDF of the $\lambda_{\text{max}}$~\cite{Satya:14}

\bea
P(\lambda_{\text{max}}=w,~N) \approx \left\{
     \begin{array} {rl}
     {\rm \exp~\Big[-\beta N^{2}\Phi_{-}}(w)\Big], & ~~~w<\sqrt{2}~\text{and}~|w-\sqrt{2}|\sim \mathcal{O}(1)\\
         \\
      {\rm \sqrt{2}N^{2/3}\mathcal{F}_{\beta}^{\prime}\Big[\sqrt{2}N^{2/3}}(w-\sqrt{2})\Big], & ~~~|w-\sqrt{2}|\sim \mathcal{O}(N^{-2/3})\\
          \\
     {\rm \exp~\Big[-\beta N^{2}\Phi_{+}}(w)\Big], & ~~~w>\sqrt{2}~\text{and}~|w-\sqrt{2}|\sim \mathcal{O}(1)\\
     \end{array} \right.
\eea
The corresponding large deviation functions as one approaches the 
critical point from below and above are respectively given by
\bea
\Phi_{-}(w)&\sim& \f{1}{6\sqrt{2}}(\sqrt{2}-w)^{3}~,~~~w\to~\sqrt{2} \\
\Phi_{+}(w)&\sim& \f{2^{7/4}}{3}(w-\sqrt{2})^{3/2},~~w\to~\sqrt{2}
\eea
We refer to the review \cite{Satya:14} and the references therein for more details.

\section{Summary and Conclusion}
To conclude, we have made a brief overview 
on the subject of extreme value statistics. We have seen that 
for uncorrelated or weakly correlated random variables, one
has a fairly good understanding of the distribution of extremum
and their limiting laws:
there exists essentially three limiting classes named
\textbf{Fr\'{e}chet,~Gumbel,~Weibull} respectively.
On the other hand, there are very few exact results known for the 
{\em strongly} correlated random variables. In these lectures,
we have discussed a few of them, but were not able to cover all
of them. 
Most of the theoretical efforts are focussed in finding
more and more exactly solvable cases which may shed some
light on the issue of the universality classes of EVS for
strongly correlated random variables. Identifying the
universality classes (if they exist) for strongly correlated
random variables is thus a very challenging and outstanding 
open problem.

Apart from the issue of the universality of the distribution of extreme values for
strongly correlated variables, there are other important questions related to extremes
that have been studied extensively in the recent past. These include the statistics of 
the time at which the extremes occur in a given time-series, statistics of record values
and their ages in a time-series and many other such questions, with applications
in finance, sports, climates, ecology, disordered systems, all the way to evolutionary biology.
Unfortunately, all these interesting topics can not be covered in these two lectures.
For some of these aspects, we refer the readers to a few recent 
reviews (from the physics literature) on extremes and related subjects and their 
applications~\cite{Krug07,satya_review10,F10,MCR10,SM13,BMS13,Wergen13,Satya:14}.


\begin{thebibliography}{9}

\bibitem{FT28} R.A. Fisher and L.H.C. Tippett,  Proc. Cambridge Phil. Soc. {\bf 24}, 180–190 (1928).

\bibitem{Gumbel} E.J. Gumbel, {\em Statistics of Extremes} (Dover, New York, 1958).

\bibitem{Gnedenko} B. V. Gnedenko, Annals of Mathematics {\bf 44}, 423–453 (1943).

\bibitem{Leadbetter} M.R. Leadbetter, G. Lindgren, and H. Rootzen, {\em Extremes and related properties of random sequences and 
processes}(Springer-Verlag, New York, 1982).

\bibitem{SM13}
G. Schehr and S.N. Majumdar, ``Exact record and order statistics of random walks 
via first-passage ideas", a book chapter in {\em  "First-Passage Phenomena and 
Their Applications", Eds. R. Metzler, G. Oshanin, S. Redner. 
World Scientific (2013)}, also available on arXiv: 1305:0639 

\bibitem{Derrida:81}
B. Derrida, Phys. Rev. B {\bf 24}, 2613 (1981).


\bibitem{Bouchaud:97}
J.-P. Bouchaud and M. Mezard, J. Phys. A: Math. Gen. {\bf 30}, 7997 (1997).

\bibitem{KM00} P.L. Krapivsky and S.N. Majumdar, Phys. Rev. Lett. {\bf 85}, 5492 (2000).


\bibitem{Satya:00}
S.N. Majumdar and P.L. Krapivsky, Phys. Rev. E {\bf 62}, 7735 (2000).

\bibitem{Dean:01}
D.S. Dean and S.N. Majumdar, Phys. Rev. E {\bf 64}, 046121 (2001).

\bibitem{ADGR01} T. Antal, M. Droz, G. Gyorgyi, and Z. Racz, Phys. Rev. Lett. {\bf 87}, 240601 (2001).


\bibitem{Raychowdhury}
S. Raychaudhuri, M. Cranston, C. Przybla and Y. Shapir, 
Phys. Rev. Lett. 87, 136101 (2001).


\bibitem{Satya:02}
S.N. Majumdar and P.L. Krapivsky, Phys. Rev. E {\bf 65}, 036127 (2002).


\bibitem{Racz:03}
G. Gyorgyi, P.C.W. Holdsworth, B. Portelli and Z. Racz, PHys. Rev. E {\bf 68}, 056116 (2003).


\bibitem{Ledoussal-Monthus}
P. Le Doussal and C. Monthus, Physica A
{\bf 317}, 140
(2003).

\bibitem{MK03} S.N. Majumdar and P.L. Krapivsky, Phyica A {\bf 318}, 161 (2003).


\bibitem{Satya:05}
S.N. Majumdar, D.S. Dean, P.L. Krapivsky, Pramana, Volume 64, Issue 6, 1175-1189, June 2005,
also available on arXiv:cond-mat/0410498

\bibitem{Satya:04}
S.N. Majumdar and A. Comtet,  Phys. Rev. Lett. 92, 225501 (2004).

\bibitem{airy}
S.N. Majumdar and A. Comtet, J. Stat. Phys. 119, 777 (2005).

\bibitem{KM05} M.J. Kearney and S.N. Majumdar, J. Phys. A: Math. Gen. {\bf 38, 4097 (2005)}.

\bibitem{Bertin-Clusel} E. Bertin and M. clusel, J. Phys. A: Math. Gen. {\bf 39}, 7607 (2006).

\bibitem{GMOR07} G. Gyorgyi, N. R. Moloney, K. Ozogany, Z. Racz, Phys. Rev. E {\bf 75}, 021123 (2007).

\bibitem{Sanjib} S. Sabhapandit and S.N. Majumdar, Phys. Rev. Lett. {\bf 98}, 140201 (2007).

\bibitem{BM07} I. Bena and S.N. Majumdar, Phys. Rev. E {\bf 75}, 051103 (2007).

\bibitem{Krug07} J. Krug, {\em Records in a changing world}, J. Stat. Mech. P07001 (2007).

\bibitem{RM07} J. Randon-Furling and S.N. Majumdar, J. Stat. Mech. P10008 (2007).

\bibitem{CS07} C. Sire, Phys. Rev. Lett. {\bf 98}, 020601 (2007).

\bibitem{CS107} C. Sire, J. Stat. Mech. P08013 (2007).

\bibitem{Burkhardt} T. W. Burkhardt, G. Gyorgyi, N. R. Moloney, and Z. Racz, Phys. Rev. E {\bf 76}, 041119 
(2007). 

\bibitem{Satya:08}
M.R. Evans and S.N. Majumdar,  J. Stat. Mech. P05004 (2008).

\bibitem{MB08} S.N. Majumdar, J.-P. Bouchaud, Quant. Fin. {\bf 8}, 753 (2008).

\bibitem{MRKY08} S.N. Majumdar, J. Randon-Furling, M.J. Kearney, and M. Yor, J. Phys. A: Math. Theor. {\bf  41}, 
365005 (2008). 

\bibitem{MZ08} S.N. Majumdar and R.M. Ziff, Phys. Rev. Lett. {\bf 101}, 050601 (2008).

\bibitem{GL08} C. Godreche and J.M. Luck, J. Stat. Mech. P11006 (2008). 

\bibitem{SMCR08} G. Schehr, S.N. Majumdar, A. Comtet, and J. Randon-Furling, Phys. Rev. Lett. {\bf 101}, 150601 (2008).

\bibitem{RMC09} J. Randon-Furling, S.N. Majumdar, and A. Comtet, Phys. Rev. Lett. {\bf 103}, 140602 (2009).

\bibitem{LW09} P. Le Doussal and K. J. Wiese, Phys. Rev. E {\bf 79} 051105 (2009). 

\bibitem{GMS09} C. Godreche, S.N. Majumdar, and G. Schehr, Phys. Rev. Lett. {\bf 102}, 240602 (2009).

\bibitem{MCR10} S.N. Majumdar, A. Comtet, and J. Randon-Furling, {\em Random convex hulls and extreme 
value statistics}, J. Stat. Phys. {\bf 138}, 955 (2010). 

\bibitem{MRZ10} S.N. Majumdar, A. Rosso, and A. Zoia, Phys. Rev. Lett. {\bf 104},  020602 (2010).

\bibitem{MRZ110} S.N. Majumdar, A. Rosso, and A. Zoia, J. Phys. A: Math. Theor. 43, 115001 (2010).

\bibitem{SL10} G. Schehr and P. Le Doussal, J. Stat. Mech. P01009 (2010).

\bibitem{NK11} J. Neidhart and J. Krug, Phys. Rev. Lett. {\bf 107}, 178102 (2011).

\bibitem{RS11} J. Rambeau and G. Schehr, Europhys. Lett. {\bf 91}, 60006 (2010); Phys. Rev. E {\bf 83}, 061146 (2011).

\bibitem{FMS11} P.J. Forrester, S.N. Majumdar, and G. Schehr, Nucl.Phys. B {\bf 844}, 500 (2011).

\bibitem{S12} G. Schehr, J. Stat. Phys. {\bf 149}, 385 (2012).

\bibitem{SMCF13} G. Schehr, S.N. Majumdar, A. Comtet, and P.J. Forrester,  J. Stat. Phys. {\bf 150}, 491 (2013).

\bibitem{DMRZ13} E. Dumonteil, S.~N. Majumdar, A. Rosso, and A. Zoia,
PNAS, {\bf 110}, 4239 (2013).
 
\bibitem{Wergen13} G. Wergen, {\em Records in stochastic processes -- Theory and applications},
J. Phys. A: Math. Theo., {\bf 46}, 223001 (2013). 

\bibitem{GMS14}  C. Godreche, S.N. Majumdar, and G. Schehr, J. Phys. A: Math. Theor. {\bf 47}, 255001 (2014).

\bibitem{FB08} Y.V. Fyodorov and J.-P. Bouchaud, J. Phys. A: Math. Theor. {\bf 41}, 372001 (2008).

\bibitem{F09}  Y.V. Fyodorov, J. Stat. Mech. P07002 (2009).

\bibitem{FLR09}  Y.V. Fyodorov, P. Le doussal, and A. Rosso, J. Stat. Mech. P10005 (2009).

\bibitem{F10} Y.V. Fyodorov, {\em Multifractality and Freezing Phenomena in Random Energy Landscapes: an Introduction},
Physica A {\bf 389}, 4229 (2010).

\bibitem{SM06} G. Schehr and S.N. Majumdar, Phys. Rev. E {\bf 73}, 056103 (2006).

\bibitem{RS09} J. Rambeau and G. Schehr, J. Stat. Mech., P09004 (2009).

\bibitem{RBKS11} J. Rambeau, S. Bustingorry, A.B. Kolton, and G. Schehr, Phys. Rev. E 84, 041131 (2011). 

\bibitem{TW94} C. A. Tracy, H. Widom, Commun. Math. Phys.
159, 151 (1994).

\bibitem{TW96} C. A. Tracy, H. Widom, Commun. Math. Phys.
177, 727 (1996).

\bibitem{DM06} D.S. Dean and S.N. Majumdar, Phys. Rev. Lett. {\bf 97}, 160201 (2006).

\bibitem{DM08} D.S. Dean and S.N. Majumdar, Phys. Rev. E {\bf 77}, 041108 (2008).

\bibitem{MV09} S.N. Majumdar, M. Vergassola, Phys. Rev. Lett.
{\bf 102}, 060601 (2009).

\bibitem{Satya:14}
S.N. Majumdar and G. Schehr, J. Stat. Mech. P01012 (2014).

\bibitem{Feller:71}
W. Feller, {\em An Introduction to Probability Theory and Its Applications}, Vol. 1, 3rd Edition, (Wiley, New Jersey, 
1971).

\bibitem{GMOR08} G. Gyorgyi, N. R. Moloney, K. Ozogany, Z. Racz, Phys. Rev. Lett. {\bf 100}, 
210601 (2008).

\bibitem{astro} M. Taghizadeh-Popp, K. Ozogany, Z. Racz, E. Regoes, and A. S. Szalay,
Astrophys. J. {\bf 759}, 100 (2012). 

\bibitem{GMORD10} G. Gyorgyi, N. R. Moloney, K. Ozogany, Z. Racz, M. Droz, 
Phys. Rev. E {\bf 81}, 041135 (2010).

\bibitem{BG10} E. Bertin and G. Gyorgyi, J. Stat. Mech. P08022 (2010) .

\bibitem{ABA12} F. Angeletti, E. Bertin, P. Abry, J. Phys. A: Math. Theor. {\bf 45}, 115004 (2012).

\bibitem{CCEF12} I. Calvo, J.C. Cuchí, J.G. Esteve, F. Falceto, Phys. Rev. E {\bf 86}, 041109 (2012).

\bibitem{ABN} B. C. Arnold, N. Balakrishnan and H. N. Nagaraja,
{\em A first course in order
statistics} ( Wiley, New York, 1992).

\bibitem{ND} H. N. Nagaraja, H. A. David,
{\em Order statistics}
(third ed.) (Wiley, New Jersey, 2003).

\bibitem{redner} S. Redner, {\em A guide to first-passage processes} (Cambridge University Press,
Cambridge, (2001)).

\bibitem{satya_review10} S.N. Majumdar, {\em Universal first-passage properties of 
discrete-time random
walks and L\'evy flights on a line: Statistics of the global 
maximum and records},
Physica A
389, 4299–4316 (2010). 

\bibitem{BMS13} A.J. Bray, S.N. Majumdar, and G. Schehr, {\em Persistence and First-Passage Properties in 
Nonequilibrium Systems}, Adv. in Phys. {\bf 62}, 225-361 (2013).


\bibitem{CM05} A. Comtet and S.N. Majumdar, J. Stat. Mech. P06013 (2005).

\bibitem{MOR11} N. R. Moloney, K. Ozogany and Z. Racz, Phys. Rev. E {\bf 84}, 061101 (2011).

\bibitem{SM12} G. Schehr and S.N. Majumdar, Phys. Rev. Lett. {\bf 108},  040601 (2012).

\bibitem{FM12} J. Franke and S.N. Majumdar, J. Stat. Mech. P05024 (2012).

\bibitem{MMS13} S.N. Majumdar, P. Mounaix, and G. Schehr, Phys. Rev. Lett. {\bf 111}, 070601 (2013).

\bibitem{PCMS13} A. Perret, A. Comtet, S.N. Majumdar, and G. Schehr, Phys. Rev. Lett. {\bf 111}, 240601 (2013).  

\bibitem{BD1} E. Brunet and B. Derrida, Europhys. Lett. {\bf 87}, 60010 (2009).

\bibitem{BD2} E. Brunet and B. Derrida, J. Stat. Phys.
{\bf 143}, 420 (2011).

\bibitem{RMS14} K. Ramola, S.N. Majumdar, and G. Schehr, Phys. Rev. Lett. {\bf 112}, 210602 (2014).

\bibitem{HZ95} T. Halpin-Healy and Y.C. Zhang, Phys. Rep.
{\bf 254}, 215 (1995).

\bibitem{Krug97} J. Krug, Adv. Phys. {\bf 46}, 139 (1997).

\end{thebibliography}
\end{document}